# Functional surface ion traps on a 12-inch glass wafer for quantum applications

J. Tao[1, a)], J. P. Likforman[2], P. Zhao[1,3], H.Y. Li[3], T. Henner[2], Y.D. Lim[1], W. W. Seit[3], L. Guidoni[2, b)] and C.S. Tan[1, c)]

[1]*School of Electrical and Electronic Engineering, Nanyang Technological University, Singapore 639798*
[2]*Laboratoire Matériaux et Phénomènes Quantiques, Université Paris Diderot-Paris7, F-75025, Paris, France*
[3]*Institute of Microelectronics, Agency for Science, Technology and Research (A\*STAR), Singapore 117685*

We report large-scale fabrication of perfectly functional radio frequency (RF) surface ion traps on a 12-inch glass substrate with a standard CMOS-compatible backend process. Established 12-inch foundry backend process of electroplated Cu with Au finish are employed to fabricate the surface electrodes directly on the glass wafer substrate. We tested a trap by loading it with laser-cooled $^{88}$Sr$^+$ ions. The trap shows a stable operation with RF amplitude in the range 100 – 230 V at 33 MHz frequency. The ion lifetime is on the order of 30 minutes for a pressure in the vacuum chamber of $5 \times 10^{-11}$ mbar, which demonstrates an exciting potential for future implementation of quantum computing system with a standard foundry process on CMOS compatible and cost-effective platform.

In the last decades, trapped ion technologies have been rapidly developed for applications in quantum computing[1,2,3], simulation[4,5] and sensing[6,7]. Invented in 2005[8] and demonstrated for the first time in 2006[9], the surface ion trap geometry, in which all the electrodes lie in the same plane enables microfabrication flexibility for complex electrode designs[10,11], opto-electronic integration[12,13] enhancing then trap scalability and functionality. Trapped ions are among the most promising systems to realize scalable quantum computers due to their capability in precise manipulation of multiple ion qubits with high fidelity and long coherence time. Some challenges still affect the implementation of this system including the use of standard CMOS processes on a preferred substrate for future opto-electronic integration[14]. Quartz[9], sapphire[15], printed-circuit board[16] and standard silicon[17] have been demonstrated for the fabrication of surface trap chips. In this work, we report traps fabricated in large-scale using a glass wafer with a standard foundry process.

Glass has superior dielectric properties with respect to silicon and it could represent a low-cost substitutional material to quartz or sapphire for surface ion trap fabrication. Compared to high resistivity (~ 5000 Ω·cm) Si substrate with a typical loss tangent of 1.5 at 20 MHz[18], the glass substrate (Corning® SWG 8.5[19]) that we adopted for trap fabrication in this work has a loss tangent of 0.025 at 5 GHz with volume resistivity of ~ $10^{10}$ Ω·cm. The glass substrates have demonstrated better RF performances compared with standard Si[20] and their good manufacturability allowed for an implementation of through-glass-vias (TGVs) with a 100% yield of 1,000 thermal-cycles reliability[21]. This technology may improve electrical and optical routing to address the scalability issue of surface trap ions towards the electro-optic integration.

In this work the trap fabrication is based on 300-mm glass wafer platform [Fig. 1 (a)] with a standard cleanroom process. Traps adopt a symmetric 5-wire geometry[8] with a RF line width of 80 μm and a nominal ion-surface distance $d$ = 75 μm [Fig. 1 (b)]. To fabricate the trap electrodes, a Ti/Cu seed layer of ~ 200 nm thickness is physically vapor deposited on a glass substrate, followed by mask-defined electroplating Cu of ~ 3,000 nm thickness and subsequent electroplating Au of ~ 200 nm thickness as a protective layer [Fig. 1 (c)]. One process issue with Cu/Au metal electroplating process is an Au overhang structure could be possibly created during Cu seed wet etching process, which leads to high leakage current between the small-gap electrodes. A process optimization by reducing the wet etching time is employed to reduce the shortage probability[22]. By conducting I-V test between either of the RF electrodes to any adjacent DC electrodes, the measured leakage current at 200 V voltage bias is ~ $10^{-8}$ A in atmosphere (22-23°C, <60% RH). After wafer fabrication, the individual trap die is diced and packaged in a 121-pin ceramic pin grid array package (CPGA, Kyocera) with a high-vacuum compatible die attach paste and 25 μm diameter Au wire bonding to make electrical connections [Fig. 1 (d)]. Thin single layer ceramic capacitors of 820±20% pF (ATC, 116UL821M100TT) are used to filter RF pickup noise on DC electrodes[23].

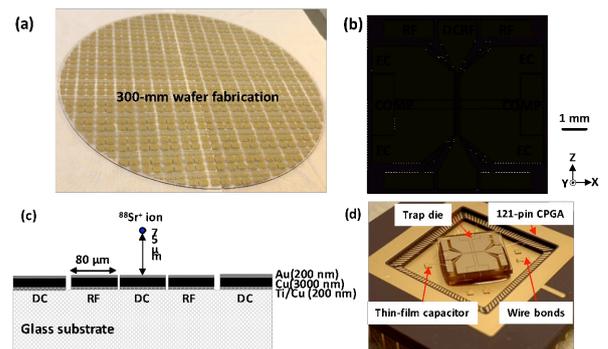

FIG. 1. (a) Trap die fabricated on a 300-mm glass wafer. (b) Schematic of trap geometry. The RF line width is 80 μm and the electrode gap is 5 μm. The name of individual electrodes are marked accordingly. (c) The cross-sectional schematic of the surface trap 5-wire geometry with RF width, ion distance and the electrode metal structure marked in the drawing. The dimensions in the schematic are not drawn to scale. (d) A trap die packaged in 121-pin CPGA with wire-bonding connections and thin-film capacitors as low-pass filters.

RF dissipation is one important issue in ion trap design. Due to the low loss tangent (0.025 at 5 GHz) of the glass substrate, it shows a low RF dissipation which simplifies the fabrication process compared with a design with electrical shielding or grounding structure to reduce the high RF loss of Si substrates[23]. We evaluated trap dissipation by conducting RF resonator tests[22] in traps with the same geometry built on high-resistivity silicon (HR-Si), HR-Si with shielding ground plane and glass substrates [Fig. 2]. The resonance curves indicate that the glass substrate traps have much higher (32 and 18 dB) resonance power peaks and improved quality factors (4.3 and 1.6 times) with respect to HR-Si and HR-Si with ground plane substrate traps respectively, which ensures a low power dissipation and high voltage step-up during trap operation.

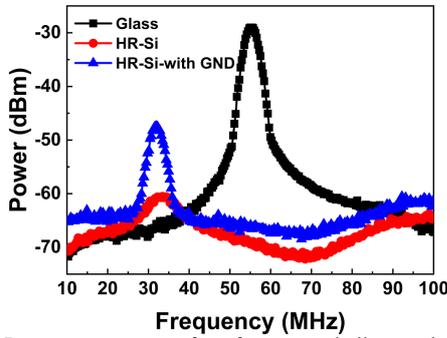

FIG. 2. Resonance curves of surface traps built on glass, high-resistivity Si and high-resistivity Si with ground substrates.

We tested the functionality of these glass substrate traps by loading one with laser-cooled $^{88}$Sr$^+$ ions. We describe in the following sections the experimental setup that we used for this purpose and the main results that we obtained.

The packaged trap is inserted in a stainless-steel ultra-high vacuum (UHV) cylindrical chamber (K.J. Lesker). A CF63 viewport of the chamber is dedicated to photon detection, three CF40 viewports to laser input and output and a CF16 viewport for lateral imaging purposes. The other CF63 flange available is used for pumping, a CF40 for electrical feedthrough (both RF and DC voltages) and two CF16 flanges are used for a Sr oven and a cold cathode vacuum gauge (Pfeiffer IKR 270) respectively. During trap operation the cell is pumped by an ion pump (IP) and a Titanium sublimator pump (TSP). We baked the newly inserted trap for a week at 130° C while pumping with a Turbo pump (Pfeiffer TC600). After the baking and the switching to IP+TSP pumping the residual pressure was in the $5\times10^{-11}$ mbar range. This demonstrates the perfect UHV compatibility of the materials used for fabrication and packaging. Adjustable DC voltages are generated by a computer-controlled DAC card (Measurement Computing PCI-DAC6703), filtered by CRCLC passive filters and then steered to FPGA via a Sub-D 9 feedthrough and shielded Kapton-insulated wires (Allectra 311-KAPM-060). RF voltage (frequency of 32.7 MHz) is supplied by a Rigol DG4162 generator, amplified in a 10 W, 50 Ohms amplifier (DeltaRF LA0005-10) and then adapted to the high impedance of the trap by a toroidal resonant transformer (step-up of 9). The resonator is directly plugged to two BNC feedthrough that are connected to the FPGA via 20 cm long Kapton-insulated coaxial cables (Allectra 311-KAP50). With this setup the maximum RF voltage amplitude at the trap is around 250 V.

An atomic beam of neutral Sr is originated by sublimation of a small Sr dendrite (Aldrich) inserted in an helicoidal tungsten filament. To create Sr$^+$ ions the atoms are photo-ionized using a two-color CW technique[24,25]. For this purpose, a 650 THz (461 nm) commercial extended cavity laser (Toptica DL Pro) addresses the $5s^2$ $^1S_0 \rightarrow 5s5p$ $^1P_1$ transition and a free running 740 THz (405 nm) diode laser that addresses the $5s5p$ $^1P_1 \rightarrow (4d^2+5p^2)$ $^1D_2$ transition that ends up in an auto-ionizing state. Both lasers are coupled in the same single mode optical fiber and then focused in the center of the trap (with a beam waist of approximately 40 μm). The power at the ion position is around 1 mW for the 650 THz and 80 μW for the 740 THz beam, respectively. In typical experimental conditions the loading-time for an ion is roughly 10 seconds. The precise frequency of the 650 THz laser is tuned with a lambda-meter (Coherent WaveMaster) and then by maximizing the fluorescence of a neutral Sr beam. No active frequency stabilization is needed during several hours.

The $^{88}$Sr$^+$ ions are Doppler-cooled addressing the $5^2S_{1/2} \rightarrow 5^2P_{1/2}$ transition (711 THz, 422 nm) with a laser beam (Toptica DL-100 laser diode). To avoid optical pumping into the metastable $4^2D_{3/2}$ state we use two additional lasers ("repumpers") addressing the 299~THz $4^2D_{3/2} \rightarrow 5^2P_{3/2}$ transition (1003 nm, Toptica DL pro laser diode) and the 290 THz $4^2D_{5/2} \rightarrow 5^2P_{3/2}$ transition (1033 nm NKT Koheras Adjustik fiber laser). With this scheme[26], it is possible to eliminate coherent population trapping issues that may appear using the repumping scheme based on the driving of the 275 THz $4^2D_{3/2} \rightarrow 5^2P_{1/2}$ transition (1092 nm)[27]. Frequency stabilization of the 711 THz cooling beam is obtained by a saturated absorption technique in a Rb cell, taking advantage of the near coincidence of the $5^2S_{1/2} \rightarrow 5^2P_{1/2}$ transition in $^{88}$Sr$^+$ with the $5^2S_{1/2} \rightarrow 6^2P_{1/2}$ transition of neutral $^{85}$Rb[28]. Infrared lasers are stabilized with a transfer lock scheme[29]. The frequency gap is filled-up and fine tuning of the cooling beam frequency is obtained using a double-pass acousto-optic modulator (AOM). The beam is coupled in a single mode optical fiber and then focused (beam waist on the order of 30 μm) at the ion position with typical power in the 2 - 50 μW range. The AOM is also used to implement a closed loop stabilization of the beam intensity by measuring a portion of the beam transmitted by the fiber on a photodiode and feeding back the signal to the AOM RF source in AM mode. Repumper lasers are both coupled in a single mode, made collinear with the cooling beam and focused at the ion position (waist on the order of 60 μm) with a typical power of 1 mW per beam.

711 THz photons scattered by the ion are collected by a home-made objective with numerical aperture of

0.4, spatially filtered in a 150 μm diameter pinhole, spectrally filtered by an interference filter (Semrock FF01-420/10) and detected by a photomultiplier in photon-counting mode (Hamamatsu H7828). The overall collection efficiency is on the order of $10^{-3}$. Alternatively, during the alignment process, it is possible to acquire images of the trapped and cooled ions (no spatial filtering) with an electron-multiplier CCD camera (Andor Luca).

Using this experimental setup, we are able to prove the functionality of the trap and to measure the ion motional frequencies. With the trap we currently studied, a small ion strings of 1 - 4 ions can be routinely loaded. For example, we acquire the stable spatially resolved fluorescence of 2-ion string [Fig. 3 (a)]. An inter-ion distance of 9.5 ± 0.5 μm is obtained by fitting the image with a double Gaussian spot [Fig. 3 (b)].

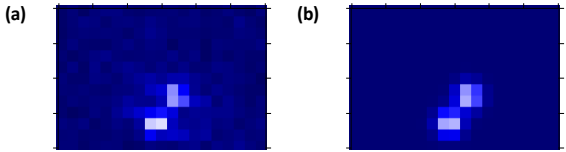

FIG. 3. (a) Fluorescence image of two trapped ions obtained for a nominal axial frequency of 300 kHz (see Table I). The image (view from above) is acquired with an electron-multiplier CCD camera. (b) Best fit of the experimental image with two Gaussian spots of the same size that gives an inter-ion distance of 9.5±0.5 μm. This distance corresponds to an axial frequency of 305±26 kHz).

We then measured the motional frequencies in the trap using a "tickle" technique[30] coupled to a sequential fluorescence acquisition. In brief an acquisition sequence is made of thousands of cycles in which first Doppler cooling is applied during roughly 1 ms, then a short (typically 10 μs) pulse of sinusoidal excitation at frequency $f$ is applied to a DC electrode, then the scattered fluorescence photons are acquired during 100 μs. During the excitation phase, the motional energy of the initially cold ion may increase in a resonant way such that the fluorescence signal acquired in the detection window is affected by Doppler shift. By scanning the frequency $f$ we measured with a precision better than 1 kHz the motional frequencies of the single ion in the trap as a function of trapping parameters (i.e. RF amplitude and DC voltages). We plot the radial frequencies measured as a function of the RF amplitudes [Fig. 4] for a given set of DC voltages (with a nominal axial frequency of 300 kHz, TABEL I). We also plot the theoretical frequencies and trap depths both calculated using an analytical model[31] as well as the frequencies simulated with a finite element method (FEM). In the experiment, the applied RF amplitudes are in the range of 120 - 160 V, the measured radial frequencies are between 3172 and 4497 kHz while the theoretical trap depth lays between 96 and 164 meV. By linear fitting the radial frequencies of measured, calculated and simulated frequencies, it shows a good matching between the theoretical and measurement data.

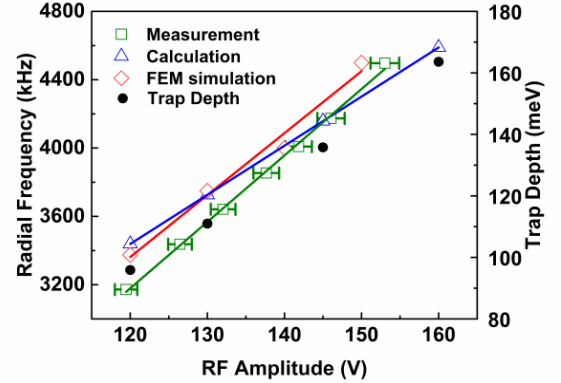

FIG. 4. Motional frequency measurement of ion radial frequency as a function of RF amplitude. The measured data are plotted and matched to analytical calculated and COMSOL simulated data.

We measured the axial frequency as a function of the nominal axial frequency [Table 1]. This nominal frequency is calculated with the analytical model[31] feed by the electrode geometry and different sets of DC voltages. There is a qualitative agreement between calculations and measurements, with a shift of roughly 100 kHz between the two. Such discrepancies were already observed in other experiments[26], and are probably due to the approximations inherent to the analytical calculation (e.g. absence of gaps between electrodes). We also used a FEM simulation by COMSOL[32] to perform the calculation [Table 1]. The FEM simulated data show a better matching to the experimental data (shift reduced to roughly 25 kHz). The improvement can be attributed to more accurate geometry (including the gaps) used to feed the FEM software.

TABEL I. DC voltage setting and the corresponding measured, nominal and COMSOL simulated axial frequencies.

| DC voltages (V) | | | Axial frequency (kHz) | | |
|---|---|---|---|---|---|
| $V_{EC}$ | $V_{COMP}$ | $V_{DCRF}$ | measured | nominal | simulated |
| +0.157 | -0.421 | -0.118 | 265 | 200 | 225 |
| +0.354 | -0.947 | -0.265 | 405 | 300 | 327 |
| +0.629 | -1.684 | -0.471 | 500 | 400 | 436 |
| +2.516 | -6.737 | -1.885 | 895 | 800 | 818 |

In this experiment, we compensated the excess of micromotion in the trap along the $x$ direction using a single photon time correlation method[33]. However, in this specific trap, the stability of such a compensation was not optimal (less than 30 min). We attribute this behavior to the contamination of the surface by point defects that may charge during time, or to some intermittent excess leak of RF voltage. Nevertheless, the lifetime of a laser-cooled ion in the trap, for a RF amplitude of 120 V is on the order of 30 min. This result is compatible with the residual pressure in the vacuum chamber.

In conclusion we demonstrated that commercially available glass Corning® SWG 8.5 is an excellent alternative to other substrates for large scale fabrication of surface ion trap devices. The first tests of a trap designed with a simple five wire geometry show a very good behavior with respect to RF dissipation, UHV compatibility and trapping properties. Further systematic studies will allow us to optimize the fabrication process to produce more reliable traps with respect to micromotion compensation, potentially based on TGV technology.

We would like to thank colleagues in Nanyang Nano-fabrication Center and technical staffs in Institute of Microelectronics, A*STAR for their technical supports on trap fabrication and packaging. The work was financially supported by A*STAR Quantum Technology for Engineering (A1685b0005). The authors would like to thank V. Tugaye and S. Guibal for fruitful discussions. This work was also partially funded by the Ile de France Region through the DIM Nano-k (DEQULOT Project).